\documentclass[aps,reprint,amsmath,amssymb,amsfonts,superscriptaddress]{revtex4-1}

\usepackage{color}
\usepackage{graphicx}   
\usepackage{array}
\usepackage{dcolumn}    
\usepackage{bm}         
\usepackage[normalem]{ulem}

\begin{document}
\title{Toroidal eigenmodes in all-dielectric metamolecules}

\author{Anna C. Tasolamprou}\email{atasolam@iesl.forth.gr}
\affiliation{Institute of Electronic Structure and Laser, FORTH, 71110, Heraklion, Crete, Greece}
\author{Odysseas Tsilipakos}
\affiliation{Institute of Electronic Structure and Laser, FORTH, 71110, Heraklion, Crete, Greece}
\author{Maria Kafesaki}
\affiliation{Institute of Electronic Structure and Laser, FORTH, 71110, Heraklion, Crete, Greece}
\affiliation{Department of Materials Science and Technology, University of Crete, 71003, Heraklion, Crete, Greece}
\author{Costas M. Soukoulis}
\affiliation{Institute of Electronic Structure and Laser, FORTH, 71110, Heraklion, Crete, Greece}
\affiliation{Ames Laboratory and Department of Physics and Astronomy, Iowa State University, Ames, Iowa, 50011, USA}
\author{Eleftherios N. Economou}
\affiliation{Institute of Electronic Structure and Laser, FORTH, 71110, Heraklion, Crete, Greece}
\affiliation{Department of Physics, University of Crete, 71003, Heraklion, Crete, Greece}

%

\date{\today}

\begin{abstract}
We present a thorough investigation of the electromagnetic resonant modes supported by systems of polaritonic rods placed at the vertices of canonical polygons. The study is conducted with rigorous finite-element eigenvalue simulations. To provide physical insight, the simulations are complemented with coupled mode theory (the analog of LCAO in molecular and solid state physics) and a lumped wire model capturing the coupling-caused reorganizations of the currents in each rod. The systems of rods, which form all-dielectric cyclic metamolecules, are found to support the unconventional toroidal dipole mode, consisting of the magnetic dipole mode in each rod. Besides the toroidal modes, the spectrally adjacent collective modes are identified. The evolution of all resonant frequencies with rod separation is examined. They are found to oscillate about the single-rod magnetic dipole resonance, a feature attributed to the leaky nature of the constituent modes. Importantly, we observe that ensembles of an odd number of rods produce larger frequency separation between the toroidal mode and its neighbor than the ones with even number of rods. This increased spectral isolation, along with the low quality factor exhibited by the toroidal mode, favors the coupling of the commonly silent toroidal dipole to the outside world, rendering the proposed structure a prime candidate for controlling the observation of toroidal excitations and their interaction with the usually present electric dipole.
\end{abstract}

\maketitle


\section{\label{sec:level1}Introduction}

The toroidal dipole, first considered by Zel'dovich in 1958, is the first of the toroidal multipoles, a peculiar electromagnetic excitation that differs from the more familiar electric and magnetic multipoles which involve the separation of negative and positive charges (the electric ones) or the closed circulation of electric currents (the magnetic ones). In contrast, the toroidal dipole results from poloidal currents circulating on a surface of a gedanken torus along its meridians. Zel'dovich connected the excitation of the static toroidal dipole, called the anapole, with parity nonconservation in  atomic spectra \cite{Zeldovich1958}, a feature also experimentally observed in later years \cite{Wood19971759,Haxton2001261}. The importance of the static anapole has been discussed for a number of solid-state systems including ferroelectric and ferromagnetic nano- and micro-structures, multiferroics, macromolecules, molecular magnets etc.  \cite{Ceulemans19981861,Popov1999330,Klaui2003,Naumov2004737,Zvezdin2009,Ungur201218554}.

In the dynamic case, an oscillating toroidal dipole emits radiation with the same angular momentum and parity properties as the electric dipole. However, the toroidal and electric dipoles have some differences: the toroidal moments interact with the time derivatives of the incident fields, the toroidal dipole radiated power scales with $\omega^6$ (rather than $\omega^4$ for the electric dipole), and their vector-potential fields do not coincide \cite{Afanasiev19954565,Dubovik1990145,Radescu19986030,Radescu2002,Gongora2006188}. Despite their distinct characteristics, toroidal multipoles are not considered in classical electrodynamics textbook \cite{Dubovik1990145,Radescu19986030,Radescu2002,Gongora2006188}. Other peculiar phenomena which have been associated with toroidal multipoles are the violation of the action-reaction equality, nonreciprocal refraction of light, and the propagation of nontrivial vector potential in the complete absence of fields \cite{Afanasiev19954565,Afanasiev2001539,Sawada2005}. In nature, materials that contain molecules of toroidal topology, such as some important macromolecules and complex proteins \cite{Hingorani200022,Simpson2000745}, are expected to exhibit toroidal-related electromagnetic properties, while anapoles have recently been connected with universe dark matter \cite{Ho2013341}.

Toroidal multipoles have attracted growing attention because of their unusual properties and their connection to the electric multipoles. However, given their silent nature, their role may easily be overshadowed by the usually much stronger electric and magnetic multipoles. Thus, special care should be exercised in systems where the relation between the time dependent charge distribution acting as the source and the far field radiation is investigated \cite{Wang199615,Wise2002317,Maier2005,Shan2006,Lal2007641,Kujala200817196}. In this respect, the rapid evolution of metamaterials has proven to be a valuable tool for understanding toroidal-related phenomena and has moreover provided the means for the direct experimental evidence of the toroidal response as seen in Ref.~\onlinecite{Kaelberer20101510}. Strong toroidal response has been observed in various systems comprising metamolecules of split-ring resonators, metallic arrays, metallic bars, etc. \cite{Ogut20125239,Dong2012,Dong201213065,Fan2013,Savinov2014,Kim2015,Watson2016}. Moreover, interesting applications have already been reported, such as a toroidal lasing spaser \cite{Huang2013} and the potential use of toroidal qubits in naturally environmentally decoupled artificial atoms \cite{Zagoskin2015}.

Recently, the range of metamaterials that support toroidal modes has been extended to all-dielectric structures \cite{Basharin2015,Miroshnichenko2015,Liu20152293}, which have the advantage of almost zero resistive losses in contrast to metallic-based toroidal metamaterials. In particular, in Ref.~\onlinecite{Basharin2015} a metamolecule of four polaritonic rods placed at the corners of a square was found to support a toroidal dipole mode. By performing scattering simulations it was shown that the toroidal mode was substantially contributing to the overall metamaterial response for a certain spectral region.

In this paper, we revisit the polaritonic-rod toroidal metamaterial. Rather than investigating the excitation of the toroidal mode through scattering simulations, we perform a comprehensive analysis of the supported eigenmodes focusing on the toroidal mode and its frequency-adjacent modes. We characterize each mode by its distinctive field distribution and by calculating the relevant multipole moments in order to identify the dominant contribution. We show that, contrary to common belief, the toroidal dipole resonance has a substantial imaginary part due almost exclusively to radiation leakage, hence favoring coupling to incoming/outgoing radiation of appropriate character, facilitating thus mode excitation/detection. We thoroughly investigate all TE$_{10}$-based collective modes supported by ensembles of $N=2-8$ polaritonic circular rods placed at the vertices of regular polygons (in TE modes the electric field is parallel to the rod axis and in particular local TE$_{10}$ modes in each cylinder constitute the building block of the toroidal mode). More specifically, we are interested in the evolution of the collective mode resonance frequencies with rod separation and particularly the spectral isolation of the toroidal mode with respect to the neighboring ones. Amongst else, we find that the cyclic metamolecule of an odd number of rods ($N=3, 5, 7$) can prove advantageous in terms of the frequency separation between the toroidal mode and its neighbors. The enhanced frequency separation, the absence of Ohmic losses and the leaky nature of the toroidal mode in the polaritonic rod metamolecules render the proposed structure a prime candidate for controlling and exploiting toroidal excitations.

The paper is organized as follows: In Sect.~\ref{sec:level2} we investigate the natural modes supported by a single polaritonic rod, focusing on the spectral range around the TE$_{10}$ (magnetic dipole) mode with resonance frequency $f_{10}$. The $N=2$ system is thoroughly examined in Sect.~\ref{sec:level3} for the purpose of understanding TE$_{10}$ collective mode formation and interpreting the evolution of the resonant frequencies with rod separation. We find that collective mode frequencies, in contrast to the LCAO experience, do not remain lower (the symmetric one) or higher (the antisymmetric one) than the single cylinder frequency $f_{10}$. Instead, they are interchanging sides depending on the rod distance. This counter-intuitive result can be explained by the leaky nature of the constituent modes. As a result, their coupling is mediated by oscillating field tails instead of evanescent ones. This explanation is quantitatively verified by substituting a linear combination of the isolated-rod modes in the frequency-squared functional of the system (corresponding to the energy functional in the case of LCAO) and minimizing it. The cross (off-diagonal) terms of the coupling matrix, responsible for frequency splitting, are indeed oscillating. Another important observation is that the oscillation of the collective mode frequencies about $f_{10}$ can be highly asymmetric leading to steep segments in the frequency-separation curve. This is because the TE$_{10}$ modes within each rod can be significantly deformed in the coupled system (compared to the isolated rod). We recover this coupling-caused current deformation with a wire model, i.e., by approximating the displacement current distribution in each rod with a pair of lumped current wires; these currents, which are determined by solving a $2N \times 2N$ eigenvalue problem, acquire asymmetric values effectively reproducing the local mode deformation. This current deformation is the analog of the dipole type charge deformation in each atomic orbital appearing in the LCAO method and being responsible for the van der Waals interactions. Finally, Sect.~\ref{sec:level4} is devoted to many-rod ($N=3-8$) systems. After a systematic analysis of the $N=3$ and $N=4$ structure (Sect. \ref{sec:level41} and \ref{sec:level42}), we proceed to compare respective systems in terms of the spectral separation between the toroidal mode and its neighbors. Systems of an odd number of rods are found to offer better spectral isolation thus favoring the excitation/detection of toroidal dipoles.

\section{\label{sec:level2}Physical System }

\begin{figure}
\includegraphics{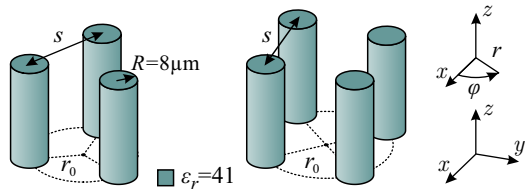}
\caption{\label{fig:rods} Schematics of the cyclic metamolecules under study for $N=3, 4$. In all cases, identical, infinitely-long polaritonic rods of circular cross-section and a radius $R=8~\mu$m are located at the vertices of regular polygons. The rod material is LiTaO$_3$ with   permittivity $\varepsilon_r=41$ around 2~THz.}
\end{figure}

The structure under study is depicted in Fig.~\ref{fig:rods} for $N=3,4$. Rods of circular cross-section are arranged at the vertices of a regular polygon lying on the $xy$-plane with their axes parallel to the $z$-axis. The cylinders extend to infinity along $z$ and possess a radius of $R=8~\mu$m. They are made of LiTaO$_3$ embedded in an infinite homogeneous medium, in this case air. LiTaO$_3$ is an ionic crystal that exhibits strong polaritonic response due to the excitation of optical phonons \cite{Huang2004543,Yannopapas2010}; LiTaO$_3$ rods can be realized with various crystal growth methods \cite{Barker19704233}. At frequencies below the phonon resonances ($\omega_T/2\pi = 26.7$ THz and $\omega_L/2\pi = 46.9$ THz is the frequency of the transverse and longitudinal phonons, respectively) LiTaO$_3$ exhibits high permittivity and very low dissipation losses.  In particular, in the frequency range under consideration, around 2~THz,  the  real part of the LiTaO$_3$  permittivity is nearly flat and equal to $\varepsilon_r=41$. Throughout this investigation the material losses have been omitted since they are negligible compared to the radiation losses for all the relevant modes. 


\begin{figure}
\includegraphics{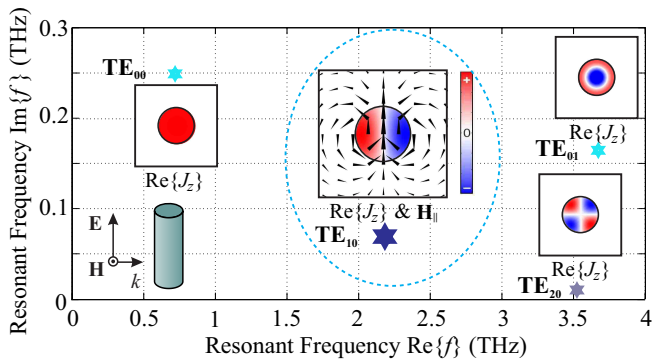}
\caption{\label{fig:singlerod} First four eigenmodes and corresponding eigenfrequencies (stars in the complex frequency plane) of TE polarization ($\mathbf{E} \equiv E_z$) for the metaatom, an isolated polaritonic rod, assuming $k_z=0$. Insets show the polarization current (color) and magnetic field (arrows in logarithmic scale)  distribution. The TE$_{10}$ (magnetic dipole) mode lies at $f_{10}$ = (2.183, 0.07) THz; the material losses for this mode are negligible ($<5\%$ of radiation losses).}
\end{figure}

The single cylindrical rod, the metaatom of the cyclic metamolecule, supports electromagnetic modes whose complex frequencies and field profiles are shown in Fig.~\ref{fig:singlerod}. These results were obtained by using the suitable Bessel (inside the cylinder) and Hankel (outside the cylinder) functions, $J_n(k_\mathrm{cyl}r)$   and   $H_n^{(1)}(k_\mathrm{air}r)$  with $k_\mathrm{cyl}=k_0n_\mathrm{LiTaO_3}$  and $k_\mathrm{air}=k_0n_\mathrm{air}$, to describe the field profile and subsequently imposing the appropriate boundary conditions at the rod interface  \cite{stratton}. A homogeneous system of linear equations is formed that admits a non-trivial solution when its determinant is zero. Assuming a wave-vector in the $xy$-plane ($k_z=0$) and TE polarization ($\mathbf{E} \equiv E_z$) the system boils down to
\begin{equation}\label{eq:natmodes}
v\frac{J'_n(u)}{J_n(u)}- u\frac{H^{(1)'}_n(v)}{H^{(1)}_n(v)} = 0,
\end{equation}
where $u=k_\mathrm{cyl}R$ and $v=k_\mathrm{air}R$. The Bessel functions are transcendental, meaning that for each value of $n$ there is an infinite number of roots denoted by the integer $m$. Therefore, $E_z$-polarized solutions are denoted by TE$_{nm}$, where subscript $n$ refers to the azimuthal and subscript $m$ to the radial order. Returning to Fig.~\ref{fig:singlerod}, the complex resonance frequencies of the supported modes in the frequency range 0-4 THz (TE$_{00}$, TE$_{10}$, TE$_{20}$ and TE$_{01}$) are shown with stars on the complex plane. The field profiles are also included: color represents the only nonzero component of the polarization current, $\mathbf{J} \equiv J_z=i\omega\varepsilon_0(\varepsilon_r-1)E_z$, while arrows represent the magnetic field which lies in the $xy$-plane. TE$_{00}$ is the lowest order mode (zero order azimuthal and radial variation) and, as an electric dipole mode, is characterized by the highest radiation losses (highest imaginary part of the resonant frequency). Next in ascending frequency, at $f_{10}$ = (2.183, 0.07)~THz, lies the TE$_{10}$ mode, which constitutes the basic element for building toroidal collective modes in the $N > 1$ systems. This mode is of magnetic dipole nature: the  current forms a loop (closes through infinity), inducing a magnetic moment which for the orientation in Fig.~\ref{fig:singlerod} (arbitrary due to cylindrical symmetry) is along the $y$ axis. The free-space wavelength at the resonance, $\lambda_{10} =137.4~\mu$m, is much larger than the radius of the rod $R=8~\mu$m ($\sim\lambda_{10}/17$); a consequence of the high rod permittivity. The quality factor is low, $Q_{10} = \Re\{f\}/2\Im\{f\} \sim 15$, indicating high radiation leakage. The field profile of the three nonzero components $\{E_z, H_r, H_\phi \}$ for $r>R$ is given by (constants aside)
\begin{equation}\label{eq:fieldprof}
\{i H_1^{(1)}(kr) \sin(\phi), \frac{H_1^{(1)}(kr)}{r}\cos(\phi),H_1^{(1)'}(kr) \sin(\phi) \}
\end{equation}
with Bessel functions taking the place of Hankel functions for $r<R$. Note the faster radial decay and the distinct azimuthal variation of $H_r$. In the spectral neighborhood of the magnetic dipole we also find the TE$_{20}$ and TE$_{01}$ modes. In parallel to the analytic solution, and having in mind the investigation of the $N > 1$ systems, we perform eigenvalue analysis with the commercial software COMSOL Multiphysics$^\circledR$ \cite{comsol} implementing the full-wave vectorial finite element method (FEM), which determines the complex eigenfrequency and field profile of each mode.

Having obtained the field distribution, we determine the dominant multipole moment for each mode. We calculate the multipole moments by integrating  the polarization currents $\mathbf{J}$  with the use of the corresponding expressions to be found in Ref.~\onlinecite{Savinov2014}. For convenience we repeat here the toroidal dipole moment expression:
\begin{equation}\label{eq:tordipole}
\mathbf{T} = \frac{1}{10c}\int d^3r  [(\mathbf{r} \cdot  \mathbf{J}) \cdot\mathbf{r}-2r^2\mathbf{J}],
\end{equation}
where $c$ is the speed of light.

The dominant multipole moments of the single cylinder eigenmodes shown in Fig.~\ref{fig:singlerod} verify their electromagnetic nature imprinted in the field distribution.  The fundamental TE$_{00}$ mode has a strong electric dipole moment component, $\mathbf{p}$,  TE$_{10}$ is characterized by a dominant magnetic dipole moment,  $\mathbf{m}$, and  TE$_{20}$ has strong magnetic quadrupole moment, $\mathbf{Q}^{(m)}$. Finally, the  TE$_{01}$ mode has a dominant toroidal dipole  moment, $\mathbf{T}$, which is intuitively expected given the formation of poloidal currents (inward and outward counter-propagating currents shown in Fig.~\ref{fig:singlerod}). Toroidal dipole excitations related to modes of the TE$_{01}$ type are discussed in Ref.~\onlinecite{Liu20152293}.

\section{\label{sec:level3}TWO-ROD SYSTEM: INTERPRETATION OF COLLECTIVE MODE EVOLUTION WITH ROD SEPARATION}

\begin{figure*}
\includegraphics{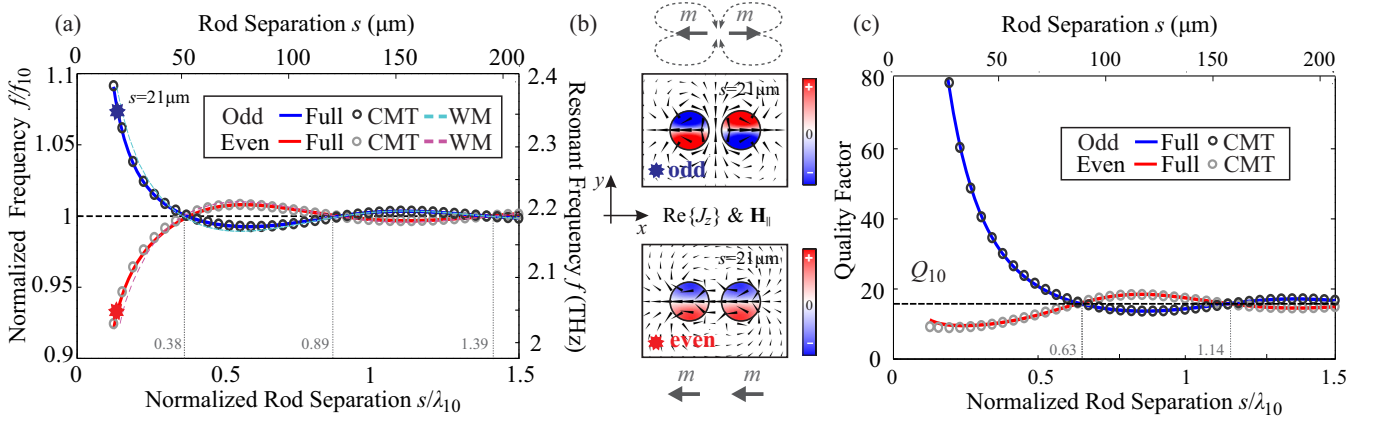}
\caption{\label{fig:tworodA} Results for $x$-oriented TE$_{10}$ collective modes of two-rod structure: (a) Evolution of resonant frequencies with rod separation $s$. The oscillation about $f_{10}$ is symmetric. The full-wave numerical solution can be accurately recovered with the coupled mode theory (CMT) and reproduced with the wire model (WM). (b) Polarization current (color) and magnetic field distribution (arrows, logarithmic scale) for the two modes at $s=21~\mu$m. (c) Quality factor evolution with rod separation.}
\end{figure*}
The TE$_{10}$ mode supported by a single polaritonic rod is the building block for the formation of the toroidal mode in Ref.~\onlinecite{Basharin2015}. Obviously, ensembles of any number of polaritonic rods in a regular polygon arrangement can also foster toroidal modes. We begin by investigating the TE$_{10}$ collective modes in the simplest case of the two-rod system, $N=2$. This way we can focus on understanding and physically interpreting the evolution of collective mode frequencies with separation distance. To this end, we complement the FEM simulations with a coupled mode theory (CMT) approach and a lumped wire model (WM), providing valuable physical insight.

The two-rod system supports four TE$_{10}$ collective modes, two consisting of $x$-oriented (i.e. along the line connecting the centers of the two rods) local TE$_{10}$ modes and two consisting of $y$-oriented local TE$_{10}$ modes (mode orientation is associated with the direction of the magnetic dipole moment). We first focus on the $x$-oriented collective modes: Fig.~\ref{fig:tworodA}(a) depicts the evolution of the resonant frequencies with rod separation for both even and odd collective modes. The same is done in Fig.~\ref{fig:tworodA}(c) for the quality factor. The field profiles of the two collective modes are depicted in Fig. 3(b) for a structure with rod separation $s=21~\mu$m. Clearly, the mode in the upper panel is odd (antisymmetric) with respect to the $yz$ mirror plane of the structure, whereas the mode in the lower panel is even (symmetric). Note that the even collective mode resembles a magnetic dipole with a net moment along the $x$ axis, whereas the odd collective mode has a zero net dipole moment, but strong quadrupole moment.

\begin{figure*}
\includegraphics{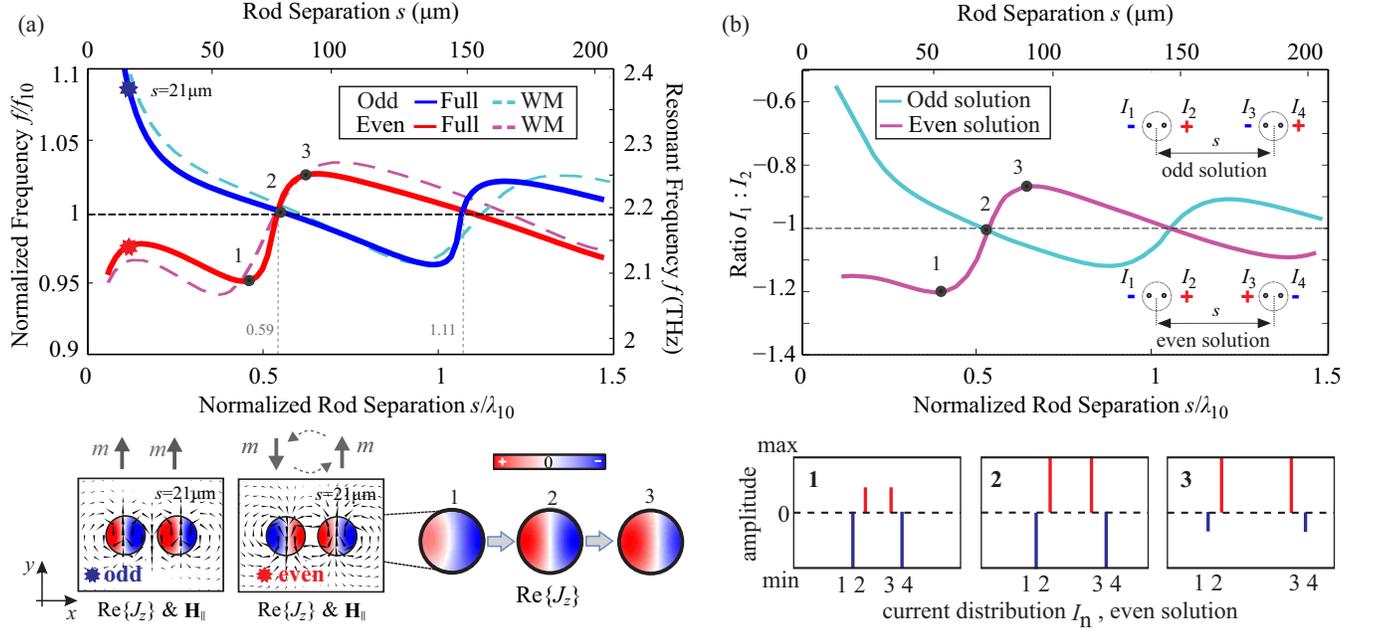}
\caption{\label{fig:tworodB} Results for $y$-oriented TE$_{10}$ collective modes of two-rod structure: (a) Evolution of frequencies with rod separation $s$. The field profiles are shown as insets for a rod separation of $21~\mu$m (color: $J_z$, arrows: magnetic field). The oscillation about $f_{10}$ is highly asymmetric. This is attributed to current redistribution within each rod due to coupling as shown in the inset depicting the distribution of the polarization current at points 1, 2 and 3 marked on the red curve. All features observed can be successfully reproduced with a lumped wire model (see text for details), the results of which are shown with dashed curves in  Fig.~\ref{fig:tworodB}(a). (b)~Ratio of currents $I_1:I_2$ [see inset for definition] determined by the four wire model as a function of $s$. The current ratio oscillates about unity in an asymmetric fashion similar to that in (a). The current values for points $1-3$ on the even solution are included as insets.}
\end{figure*}

As anticipated, coupling results in frequency splitting, i.e., two collective modes with frequencies above and below the isolated-rod frequency $f_{10}$, respectively. What is interesting is that the odd (even) mode does not remain strictly above (below) $f_{10}$. Rather, the resonant frequencies oscillate (in this case symmetrically) about $f_{10}$. In fact, the shape of the oscillation can be described quite accurately by $Y_1(2 \pi s/ \lambda_{10} )/ s $. This is consistent with Ref.~\onlinecite{Fan:1999}: the propagating state mediating the coupling in our case is the radiation leakage of the modes themselves, described by Eq.~\eqref{eq:fieldprof}. Since it is only $H_r$   that is nonzero along the coupling direction for this specific local dipole orientation, a translation operation along $x$ primarily results in a scaling of the resonator coupling coefficient with $H_1^{(1)}(kr)/r$. Naturally, the imaginary part of this coupling coefficient can be associated with the collective mode resonant frequency (whereas the real part with $\Im\{f\}$), explaining the  $Y_1(2 \pi s/ \lambda_{10} )/ s $ variation. The intersections $f_e=f_o$ ($e$ for even, $o$ for odd) where frequency splitting vanishes are clearly marked in Fig.~\ref{fig:tworodA}(a). They are seen to nicely correspond to the zeros of the $Y_1(2 \pi s/ \lambda_{10}) $  function ($s/ \lambda_{10}  = 0.35, 0.86, 1.37$). Note that intersections occur at $f_{10}$, i.e., it holds $f_e=f_o=f_{10}$. Finally, the decay of the oscillation is physically anticipated, since power density decreases with separation and, thus, coupling becomes weaker. There is also quality factor splitting, Fig.~\ref{fig:tworodA}(d), originating from the fact that isolated modes couple in the far field as well, leading to constructive or destructive interference of the radiated fields \cite{Gentry:2014}. The shape of the $\Im\{f\}$ oscillation (not shown) can be described by $J_1(2 \pi s/ \lambda_{10})/s $. This is manifested in the quality factor by the intersections $Q_e=Q_o$ which occur at the zeros of the $J_1(2 \pi s/ \lambda_{10}) $ function $(s/ \lambda_{10} = 0.61, 1.12)$. Again, it holds $Q_e=Q_o=Q_{10} \sim 15$.

The full-wave results can be accurately reproduced with a CMT framework \cite{Haus:1991,Popovic:2006} which amounts to substituting a linear combination of the isolated rod modes in the frequency-squared functional of the two-rod system and minimizing (in direct analogy with the LCAO method). Details regarding the formulation can be found in Appendix~\ref{sec:CMT}. The results are shown in Fig.~\ref{fig:tworodA}(a),(c) with circular markers. Clearly, the agreement with the full-wave simulations of the coupled system is exceptionally good corroborating the validity of the results.

We now return to the oscillations of the collective mode frequencies about $f_{10}$ with increasing $s$, which is an atypical and initially counter-intuitive result. It can be explained by the fact that the two isolated-rod modes forming the collective mode are leaky. As a result, their coupling is mediated by oscillating field tails instead of evanescent ones (which is the case for bound waveguide modes in electromagnetics or wavefunctions in quantum mechanics). This claim can be further corroborated by turning to CMT. More specifically, the cross term of the coupling matrix which is responsible for frequency splitting (see Appendix~\ref{sec:CMT}) acquires positive or negative values depending on rod separation. Being an overlap integral of the two isolated-rod mode profiles over one rod's cross-section, this is only possible when oscillating mode tails are involved, not evanescent ones. As mentioned, the periodic oscillation of $f_e$ and $f_o$ about $f_{10}$ with a shape determined by the propagating state mediating resonator coupling has been also noted in the context of guided-wave photonic circuits \cite{Fan:1999}.

The collective modes of $y$-oriented TE$_{10}$ local modes are examined in Fig.~\ref{fig:tworodB}. The field distribution of both odd and even modes for a separation distance $s=21~\mu$m are presented as insets. The even collective mode, red line in Fig.~\ref{fig:tworodB}(a), is characterized by the presence of polarization currents that oscillate in the inward and outward parts of each rod with opposite directions. These currents produce a vortex of the magnetic field that threads both current loops and correspond to a precursor of the toroidal dipole mode  which will be thoroughly discussed in Sect.~\ref{sec:level4}. In Fig.~\ref{fig:tworodB}(a) we observe that unlike the $x$-orientation case, the oscillations about $f_{10}$  are highly asymmetric, of larger amplitude, and with steep transitions between the local minima and maxima for both even and odd collective modes. This behavior can be explained as follows: In the case of $y$-orientation the maximum of the radiation pattern is towards the adjacent rod. This leads to the deformation of the polarization current distribution within the rods, significantly affecting the resonant frequencies of the collective modes. The phenomenon is analogous to the charge redistribution within each atom which leads to induced dipole moments and the van der Waals interaction. An inset in Fig.~\ref{fig:tworodB}(a) presents the distribution of the polarization current in the rods at points 1, 2 and 3 marked along the red curve (given the symmetry of the fields, only one rod is presented). The local dipoles in the coupled system are most significantly deformed at points 1 and 3 where the resonant frequency is farthest away from $f_{10}$. In contrast, at point 2 where $f=f_{10}$ the local dipole modes are almost perfectly symmetric.

Although this highly asymmetric oscillation cannot be described with a closed-form function as in the $x$-orientation, the intersections $f_e=f_o$ still correspond to the zeros of $J_1(2 \pi s/ \lambda_{10})$, $s/ \lambda_{10} = 0.61, 1.12$,  as one would anticipate given that it is now $E_z$ that mainly mediates resonator coupling, see Eq.~\eqref{eq:fieldprof}. This time, the collective mode frequencies at the intersections are not exactly equal to $f_{10}$. At the intersection points the current distribution in each rod is almost, but not exactly, symmetric. This small asymmetry, as opposed to the perfect symmetry in an isolated rod, accounts for the small difference between $f_e=f_o$ and $f_{10}$. In other words, the self-effect known as coupling induced frequency shift (CIFS) \cite{Popovic:2006}, quantified by the main diagonal elements of the coupling matrix in the CMT framework is not as weak and becomes noticeable when frequency splitting vanishes.

CMT is not capable of accurately recovering the collective mode frequencies in this case, since it cannot account for current deformation: the collective modes are built directly from the isolated, perfectly symmetric TE$_{10}$ modes. However, the features observed in Fig.~\ref{fig:tworodB}(a) can be successfully reproduced by making use of a simple lumped wire model (see Appendix~\ref{sec:WM} for details), capable of reproducing the fact that up and down currents in each rod are not in general equal. More specifically,  the current distribution in each rod is described with a pair of current-carrying thin wires as shown in the schematic of Fig.~\ref{fig:tworodB}(b). The distance between the two wires within each rod was kept constant at $7~\mu$m as it has been found to effectively reproduce the TE$_{10}$ modes under consideration. Having two degrees of freedom for each rod we can effectively allow for mode deformation. For the two-rod structure, a $4 \times 4$ eigenvalue problem is formulated, which can be solved to produce four eigenvalues, $Z$, and four eigenvectors, $\mathbf{I}$, for each value of separation distance $s$. The eigenvalues $Z$ correspond to the collective impedance in each wire and the eigenvectors $\mathbf{I}$ to the currents in the wires. The imaginary part of the impedance is proportional to the resonant frequency of the system (the real part accounts for losses and can thus be associated with the imaginary part of the resonant frequency), whereas the current vector describes current redistribution. Two of the four solutions obtained correspond to the $y$-oriented odd and even collective modes. The imaginary part of the normalized eigenvalues $Z/Z_{10}$ is plotted in Fig.~\ref{fig:tworodB}(a) with dashed lines. The eigenvalue $Z_{10}$ corresponds to a single pair of wires at a distance equal to $7~\mu$m, i.e., the isolated-rod TE$_{10}$ mode. The results are found to successfully reproduce the features of the collective mode frequency evolution.
	
The deformation of the local dipoles is manifested in the ratio $I_1:I_2$ (or, equivalently, $I_3:I_4$) of the further-away currents to the nearby currents [see insets in Fig.~\ref{fig:tworodB}(b)]. This ratio is plotted versus the separation distance, $s$, in Fig.~\ref{fig:tworodB}(b). Just like the collective mode frequencies, it oscillates around unity in a similarly asymmetric fashion. The amplitudes of the four currents for points 1, 2 and 3 on the even branch are also included as insets. Just like the eigenmode polarization currents [insets in Fig.~\ref{fig:tworodB}(a)] the lumped wire currents are most significantly deformed at points 1, 3 where the imaginary part of the normalized eigenvalue $Z/Z_{10}$ is furthest away from unity. In contrast, at point 2 where $\Im\{Z\}\sim\Im\{Z_{10}\} $ the inside and outside currents are equal in amplitude.

We conclude that when collective modes of leaky resonant modes are concerned, irrespective of the symmetry (even or odd with respect to the mirror planes of the structure), each collective mode can be found on either side of $f_{10}$ depending on the rod distance. In addition, depending on the radiation pattern the fields of each resonator can significantly disturb each other leading to asymmetric oscillations of the resonant frequencies about $f_{10}$ with large amplitudes.

\section{\label{sec:level4} MANY-ROD CYCLIC METAMOLECULES}

We turn now to the study of many-rod cyclic metamolecules. In each case we solve for the TE$_{10}$-based collective modes and examine their evolution with rod separation. We are particularly interested in identifying the toroidal mode supported by such systems and in determining the conditions for spectrally separating it from its neighbors.

\subsection{\label{sec:level41}Three-rod cyclic metamolecule}

\begin{figure*}
\includegraphics{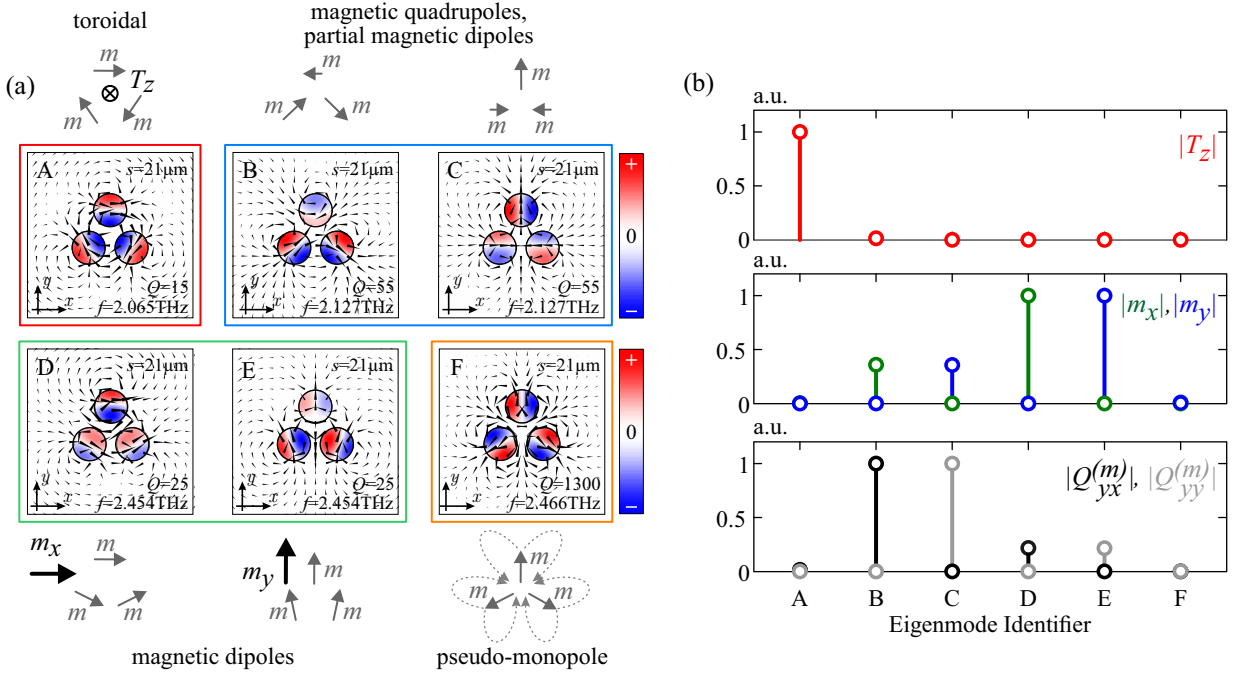}
\caption{\label{fig:threerodNew} (a)~TE$_{10}$ collective modes supported by the three-rod cyclic metamolecule with a rod separation of 21 $\mu m$ as obtained by the accurate full wave numerical calculation. Color shows the current distribution and arrows the magnetic field. They are numbered according to their resonant frequency in ascending order. (b)~Toroidal dipole moment, $T_z$, magnetic dipole moments, $m_x$ and $m_y$, and  quadrupole magnetic moments, $Q_{yx}^{(m)}$ and  $Q_{yy}^{(m)}$ (absolute values) for each of the six modes A to F of Fig.~\ref{fig:threerodNew}(a). The moments are shown normalized to the maximum value within each panel.  }
\end{figure*}

The three-rod metamolecule supports six TE$_{10}$ collective modes. Their profiles are depicted in Fig.~\ref{fig:threerodNew}(a) for a configuration with a rod separation $s=21~\mu$m $(s/\lambda_{10} \sim 0.15)$. Color corresponds to $J_z$ (the sole component of the polarization current) and arrows correspond to the magnetic field (logarithmic scale). The logarithmic scale helps to better convey the direction of the magnetic field. The collective modes are ordered according to their resonant frequency (real part) in ascending order. Modes B and C and modes D and E, respectively, have the same resonant frequency, i.e., they are degenerate. Figure \ref{fig:threerodNew}(b) presents the calculated absolute values of the relevant  multipole  moments and in particular the toroidal dipole moment, $T_z$, the magnetic dipole moments, $m_x$ and $m_y$, and quadrupole magnetic moments, $Q_{yx}^{(m)}$ and  $Q_{yy}^{(m)}$, for each of the six collective modes. The moments are  calculated by integrating  each eigenmode current distribution using the suitable formulas found in Ref.~\onlinecite{Savinov2014}. In order to provide a  fair comparison between the collective modes, we normalize each current distribution with the square root of the stored  electric energy in the corresponding eigenmode.

Clearly the fields in each rod closely resemble the TE$_{10}$ mode of the single rod (properly rotated depending on the specific mode), as one would anticipate for TE$_{10}$-based collective modes. Note that all collective modes in Fig.~\ref{fig:threerodNew}(a) involve strong interaction between the constituent modes as evidenced by the directions of the respective $E$-field maxima. In fact, these interactions result in the deformation of the local currents, as discussed in Sect. \ref{sec:level3}. Depending on the coupling strength, the local current deformation may give rise to a nonzero total current in each rod and hence to a nonzero electric dipole moment, which would otherwise be zero due to the perfectly antisymmetric current distribution in the TE$_{10}$ isolated-rod mode.  Mode A is the toroidal mode supported by the structure. Its distinct trademark is the ring-like structure of the magnetic field threading all three current loops. This conclusion is rigorously proved by the results shown in Fig.~\ref{fig:threerodNew}(b) according to which, mode A consists exclusively of the toroidal dipole moment (and a non-resonant electric dipole moment appearing only when the net current in the metamolecule is nonzero). Notice that the coexistence of both toroidal and electric dipole moments provides the possibility of mutual cancellation of their fields outside the source region (by adjusting their magnitude and phase) and therefore of extremely high $Q$-factors. The current distribution of modes B, C corresponds to magnetic quadrupole  moments, an observation verified by the dominant $Q_{yx}^{(m)}$ and  $Q_{yy}^{(m)}$ values. Moreover, the three partially cancelling each other magnetic dipole moments of modes B, C produce nonzero net dipole magnetic moments (directed along $x$ and $y$, respectively) also imprinted in the nonzero values of $m_x$ and $m_y$; for this reason along with the term quadrupole we also use the term partial magnetic dipole. In modes D and E, all three local magnetic dipole moments combine in a net moment, $\mathbf{m}$, which is parallel to the $x$ or $y$ axis, respectively, also proven by the high $m_x$ and $m_y$ values. Finally, in mode F the values of the above moments are insignificant. Mode F seems to radiate its magnetic field radially (the magnetic lines of course return back). Based on this phenomenological observation and in order to preserve a consistency in the terminology of all the $N$-rod systems under consideration, we term this type of mode hereinafter magnetic pseudo-monopole.  Nonetheless we mention $N$=3, the pseudo-monopole exhibits a nonzero magnetic octupole moment, $O_{yxx}^{(m)}$, also marked in the deduced radiation pattern.
Concluding the characterization of the modes, it should be stressed here that obviously in a scattering scenario, depending on the specific excitation (direction of incidence, phase front etc.) the contribution of each moment to the scattered power is expected to vary.


\begin{figure*}
\includegraphics{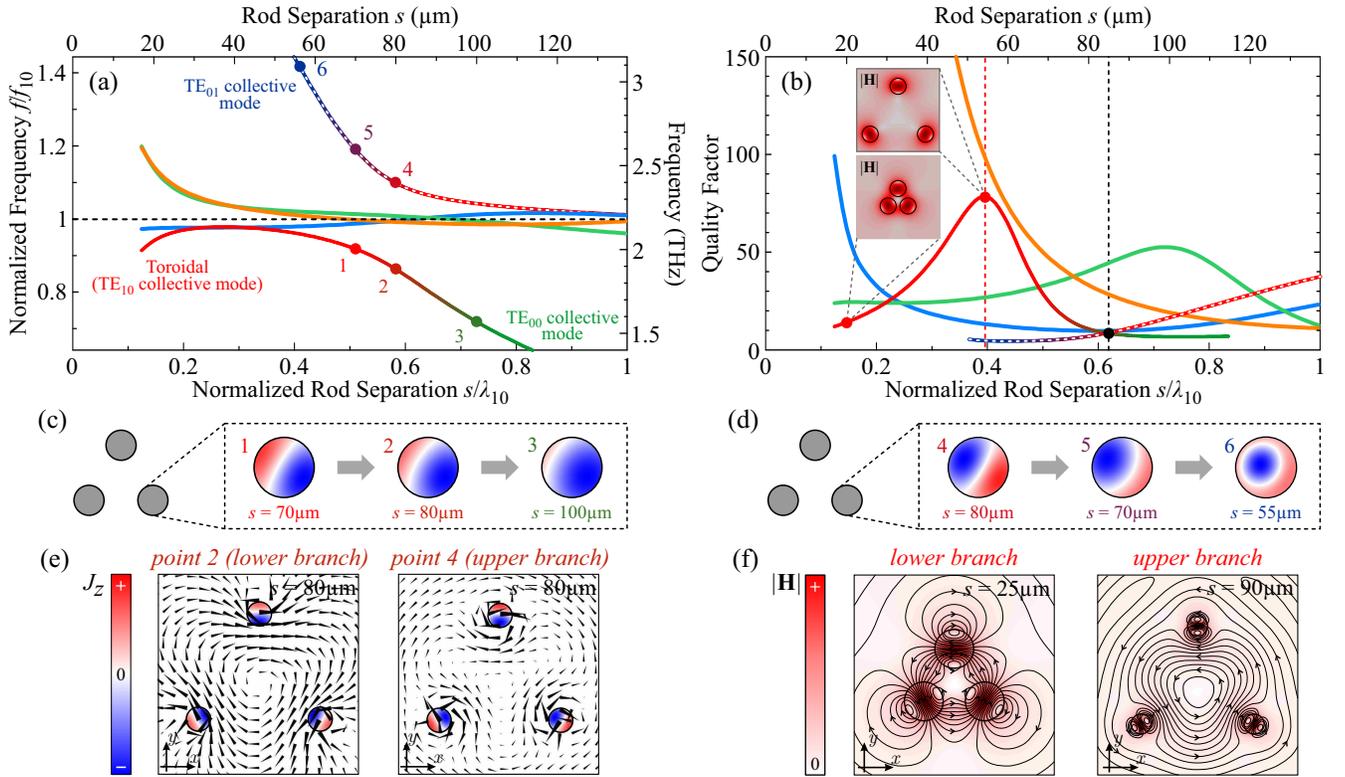}
\caption{\label{fig:threerod} (a,b)~Mode frequencies and quality factors as a function of rod separation for the six collective modes in the $N=3$ system. Curve colors correspond to the frames surrounding the mode distributions in Fig.~\ref{fig:threerodNew}(a). (c)~Evolution of the lower branch of toroidal nature into a TE$_{00}$ collective mode with increasing $s$. Points $1-3$ are clearly marked in Fig.~\ref{fig:singlerod}(a). (d)~Evolution of the upper branch (a kind of spatially diffused toroidal mode) into a TE$_{01}$ collective mode with decreasing $s$. (e)~Comparison of field distributions for points 2 and 4 ($s = 80~\mu$m) belonging to different branches of mode  A (see Fig. ~\ref{fig:threerodNew}(a)). The magnetic fields of local dipoles can connect with each other producing a stretched toroidal mode (point 2, lower branch) or close locally (point 4, upper branch). This qualitative difference is noted in (a) and (b) by using a dashed line for the upper branch. (f)~Magnetic field lines (color represents $|\mathbf{H}|$) 
in the lower ($s = 25~\mu$m) and upper ($s = 90~\mu$m) branch of mode A, respectively, emphasizing the difference between the two branches.}
\end{figure*}

The evolution of the collective mode frequencies with rod separation is depicted in Fig.~\ref{fig:threerod}(a). In agreement with the behavior of the two-rod structure, each collective mode can cross to the other side of the TE$_{10}$ resonant frequency, $f_{10}$, as rod separation increases. The unique feature in this case is that mode A [red line in Fig.~\ref{fig:threerod}(a,b)] consists of two disconnected branches. The lower branch is a true toroidal mode at small separations and evolves into a TE$_{00}$ collective mode for large separations. This can be verified by observing Fig.~\ref{fig:threerod}(c) which demonstrates that the field inside the rods is progressively deformed until the first-order azimuthal variation vanishes. The upper branch is a kind of spatially diffused toroidal mode for large $s$ [see Fig.~\ref{fig:threerod}(e,f)] and evolves into a TE$_{01}$ collective mode as separation decreases, Fig.~\ref{fig:threerod}(a). Notice in Fig.~\ref{fig:threerod}(d) how the first-order azimuthal variation gradually gives its place to a first-order radial variation. This transformation to collective modes based on other than the TE$_{10}$ single-rod mode is to be expected when the frequencies of the TE$_{10}$-based collective modes approach the frequencies of other single-rod modes. Note that in the $N=3$ structure frequency splitting is stronger compared to the two-rod structure (compare maximum deviation from $f_{10}$ in Fig.~\ref{fig:threerod} and Figs.~\ref{fig:tworodA},\ref{fig:tworodB}, respectively). In fact, considering the high imaginary part of the TE$_{00}$ and TE$_{01}$ modes (Fig.~\ref{fig:singlerod}), the corresponding collective modes are expected to deviate even more significantly from $f_{00}=0.72$ THz and $f_{01}=3.67$ THz, respectively (Fig.~\ref{fig:singlerod}). As a result, the evolution of a TE$_{10}$ collective mode into a TE$_{00}$/TE$_{01}$ collective mode becomes possible, something that was not witnessed in the $N=2$ structure. Obviously, coupling mode theory can describe this phenomenon only if the scheme includes (besides the TE$_{10}$) the TE$_{00}$ and TE$_{01}$ modes for each rod.

It is also important to note the different characteristics of the magnetic field distribution for points 2 and 4 in Fig.~\ref{fig:threerod}(a) which share the same $s$ value ($80~\mu$m) but belong to different branches of collective mode A. They are highlighted in Fig.~\ref{fig:threerod}(e). The magnetic fields of local dipoles in point 2 (lower branch) connect with each other forming a unidirectional magnetic field vortex threading the  current loops in each rod, characteristic of a toroidal mode. On the other hand, in point 4 the magnetic field forms again broad and spatially diffused closed loops which, however, avoid the rods which form local dipoles. The above observation holds for the entire upper and lower branch as illustrated in Fig.~\ref{fig:threerod}(f) where the magnetic field lines are compared for two different points on the lower ($s=25~\mu$m) and upper ($s=90~\mu$m) branch of collective mode A. In the lower branch, the magnetic field lines pass through the rods threading the  current loops, whereas on the upper branch they bypass them. To emphasize this qualitative difference the lower branch is indicated in Fig.~\ref{fig:threerod}(a,b) with a continuous line, while the upper one with a dashed line.

\begin{figure*}
\includegraphics{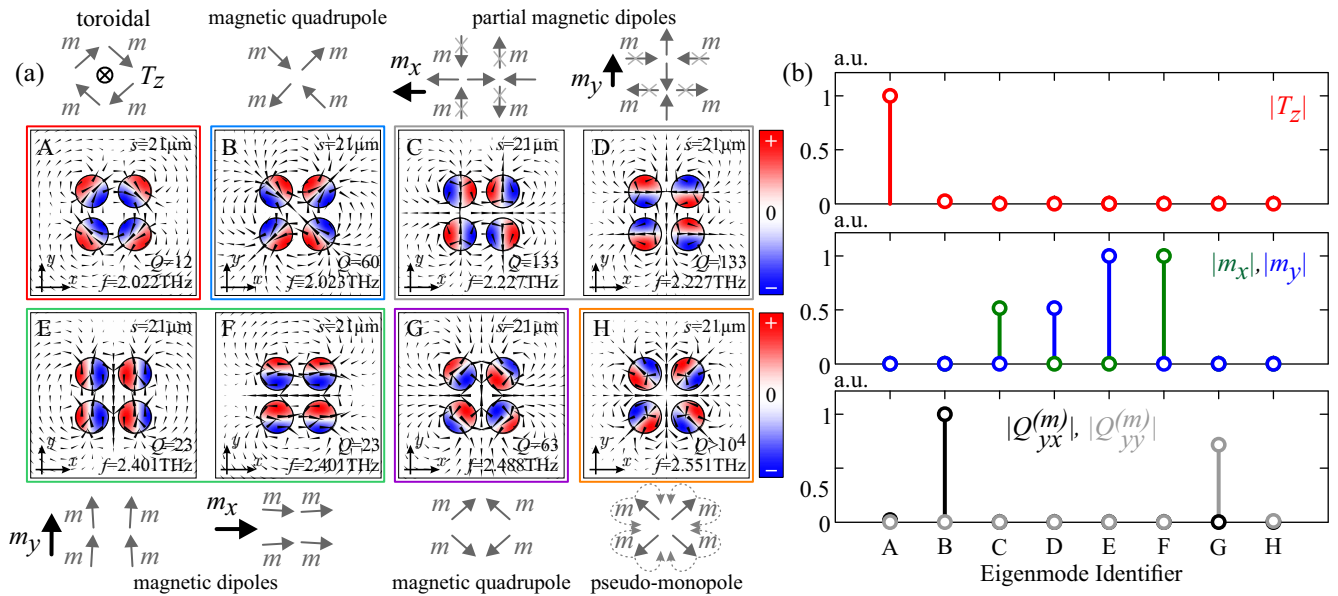}
\caption{\label{fig:fourrodNew}  (a)~TE$_{10}$ collective modes supported by the four-rod cyclic metamolecule with a rod separation of 21 $\mu m$ as obtained by the accurate full wave numerical calculation. Color shows the current distribution and arrows the magnetic field. They are numbered according to their resonant frequency in ascending order. (b)~Toroidal dipole moment, $T_z$, magnetic dipole moments, $m_x$ and $m_y$, and  quadrupole magnetic moments, $Q_{yx}^{(m)}$ and  $Q_{yy}^{(m)}$ (absolute values) for each of the six modes A to H of Fig.~\ref{fig:fourrodNew}(a). The moments are shown normalized to the maximum value within each panel. }
\end{figure*}

The evolution of the collective mode quality factors with rod separation is depicted in Fig.~\ref{fig:threerod}(b). Importantly, the toroidal mode possesses a relatively low quality factor, indicating strong coupling to plane waves which favors its excitation/detection. Especially for $s<30~\mu$m its quality factor is the lowest among all collective modes. A local maximum is observed at $s=55~\mu$m $(s/\lambda_{10} \sim 0.4)$. At this point the magnetic field torus is least connected since the magnetic field is mainly localized inside the rods. The insets in Fig.~\ref{fig:threerod}(b) demonstrate this weakening of the magnetic field torus. A second observation is that the quality factor of mode B (the neighbor of the toroidal) shown in light blue decreases with rod separation. Therefore, the linewidth of the resonance (quantified by the half power bandwidth, HPBW) increases, something that affects the spectral isolation of the toroidal mode. This will be examined in detail in Sect.~\ref{sec:level43}. Finally, although the two branches of the toroidal mode are disconnected, if we wanted to define a transition point between them this could be where the quality factor curves (red and blue) intersect. In fact, this happens at a normalized rod separation of $\sim0.62$ ($s = 85~\mu$m), consistent with the first zero of the $J_1(2 \pi s/ \lambda_{10})$ function.

\subsection{\label{sec:level42}Four-rod cyclic metamolecule}

\begin{figure}
\includegraphics{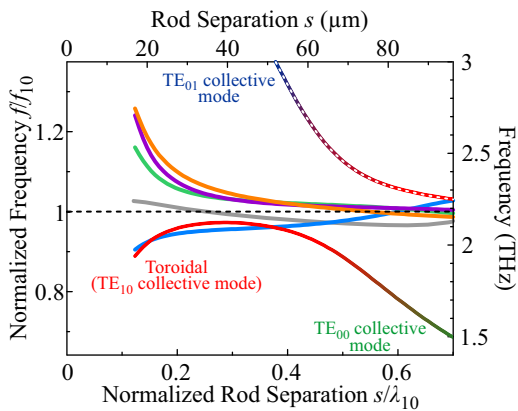}
\caption{\label{fig:fourrod}   Resonant frequency evolution with rod separation for the TE$_{10}$ collective modes in the four-rod cyclic metamolecule. Curve colors correspond to the frames surrounding the mode distributions in Fig.~\ref{fig:fourrodNew}(a).}
\end{figure}

Turning to the four-rod metamolecule we find eight TE$_{10}$ collective modes. Their field profiles are depicted in Fig.~\ref{fig:fourrodNew}(a) for a configuration with $s = 21~\mu$m. They appear according to their resonant frequency (real part) in ascending order. Note that modes C \& D and modes E \& F, respectively, are degenerate. Importantly, the neighbor of the toroidal (mode B) is not doubly degenerate as in the three-rod structure. As will be shown in detail in Sect.~\ref{sec:level43}, this is a distinct difference between even-numbered and odd-numbered systems and affects the spectral isolation of the toroidal mode.  The absolute value of the relevant multipole moments (toroidal dipole, magnetic dipole and magnetic quadrupole moment)  for each mode is presented in  Fig.~\ref{fig:fourrod}(b).  Regarding mode characteristics, mode A (the lowest-frequency TE$_{10}$ collective mode for small separations) is the toroidal mode of the structure, exhibiting a large toroidal moment, $T_z$. It also features the lowest quality factor among all collective modes for small separations, indicating strong coupling to plane waves. The excitation of the toroidal dipole in a four-rod-based metamaterial has been thoroughly discussed in Ref.~\onlinecite{Basharin2015}. The toroidal mode appears in the relevant scattering numerical experiment and at the same time its critical contribution to the multipole decomposition is demonstrated. Modes B and G are clearly magnetic quadrupoles verified both by the field distribution and by their large magnetic quadrupole moments, $Q_{yx}^{(m)}$ and $Q_{yy}^{(m)}$, respectively.  Modes C and D have  a nonzero net magnetic moment also evident in the increased values of $m_x$ and $m_y$, and are termed  partial magnetic dipoles. In modes E and F all local moments are aligned giving rise to a strong net dipole moment with a clear direction which is also reflected in the large dipole moments,    $m_x$ and $m_y$, in
Fig.~\ref{fig:fourrodNew}(b); they are, thus, termed magnetic dipoles. Finally, as in the three-rod structure, the highest-frequency collective mode for small separations (mode H) is phenomenologically termed a magnetic pseudo-monopole.

The evolution of collective mode frequencies with rod separation is depicted in Fig.~\ref{fig:fourrod}. The behavior is entirely analogous to the three-rod case [cf. Fig.~\ref{fig:threerod}(a)]. Mode A consists of two disconnected branches. The lower branch is of toroidal nature and evolves into a TE$_{00}$-based collective mode for large separations. On the other hand, the upper branch is a spatially diffused toroidal mode (indicated with a dashed line) with the magnetic field lines bypassing the rods, as in Fig.~\ref{fig:threerod}(f). It evolves into a TE$_{01}$-based collective mode as separation decreases. The main difference with the three-rod structure is that the toroidal mode is not well-separated from its neighbor [light blue line in Fig.~\ref{fig:fourrod}] for small $s$ values. The two modes start separating for $s>60~\mu$m, where mode A has already begun evolving into a TE$_{00}$-based collective mode and where the resonance linewidth of mode B is quite wide [see Fig.~\ref{fig:linewidth}(b)]. As a result, exciting the mode A without exciting mode B as well seems challenging. This indicates that the three-rod structure can provide better toroidal-dominated response compared to the one observed for the four-rod structure in Ref.~\onlinecite{Basharin2015}.

\subsection{\label{sec:level43}Toroidal dipole: Spectral isolation in $N=3-8$ cyclic metamolecules}
\begin{figure*}
\includegraphics{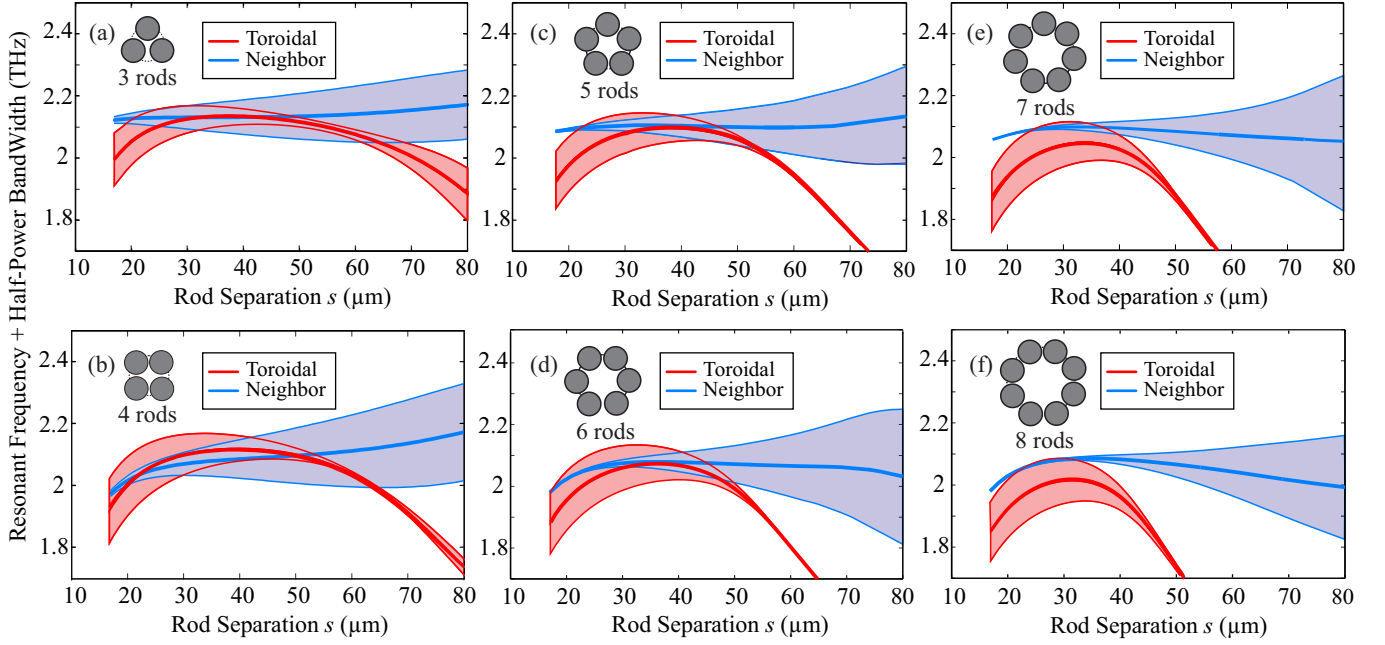}
\caption{\label{fig:linewidth} Resonant frequency and half-power bandwidth of the toroidal mode and its closest neighbor as a function of $s$ for systems of $N=3-8$ rods (a-f respectively). The optimum operating point is $s_\mathrm{min} = 17~\mu$m where the neighbor’s linewidth is at a minimum and frequency separation at a local maximum (at least before the toroidal mode begins evolving into a TE$_{00}$ collective mode). As $N$ increases spectral isolation is enhanced. Importantly, odd-numbered structures systematically provide higher spectral isolation than even-numbered systems.}
\end{figure*}

Having systematically identified the TE$_{10}$ collective modes and their various features for systems of $N=3,4$ rods, we are now interested in the structure that provides the highest degree of spectral isolation for the toroidal mode, something that is expected to facilitate its unambiguous excitation/detection. To this end, we examine structures with $N=3-8$ and compare them on this basis. We find that odd-numbered structures are advantageous (something already indicated by the three- and four-rod structures, Sects.~\ref{sec:level41} and \ref{sec:level42}). A physical interpretation for this feature is provided below. Figure~\ref{fig:linewidth} depicts the resonant frequency evolution of the toroidal mode (red curves) and its closest neighbor (blue curves) for $N = 3-8$ systems in the rod separation range $17~\mu\mathrm{m}<s<80~\mu$m. In order to investigate the spectral isolation of the toroidal mode, apart from the central frequencies we also need the resonance linewidths of the neighbor and the toroidal. Thus, we also plot the half-power bandwidth (HPBW), $\Delta f_{3dB} = f/Q$ (shaded areas). In all cases, for small rod separation values the HPBW of the toroidal mode is significant, of the order of 200GHz; for larger separation values it decreases. That is the toroidal response is expected to be wideband for small rod separation and narrowband for large rod separation. On the contrary the neighbor mode exhibits low HPBW, in the order of a few GHz, for small separation values and increases as rod separation becomes larger. This is a feature that may further contribute to the identification of the toroidal mode.

We now focus on the $N=3$ and $N=4$ case presented in Fig.~\ref{fig:linewidth}(a),(b). In the $N=3$ system, closest to the toroidal mode lies the quadrupole/partial magnetic dipole shown in Fig.~\ref{fig:threerod}(a) (mode B). As is evident in Fig.~\ref{fig:linewidth}(a), the two modes are farthest away for $s<25~\mu$m and $60~\mu\mathrm{m}<s<80~\mu$m. In the intermediate region the two resonances significantly overlap. Exploiting the $60~\mu\mathrm{m}<s<80~\mu$m region is not a favorable option since the toroidal mode has already begun evolving into a TE$_{00}$ collective mode. In addition, the HPBW of the neighboring mode is significantly increased. The optimum operating point is, thus, $s_\mathrm{min} =17~\mu$m where the quality factor of the second mode is maximum leading to a resonance span of only 20~GHz, much smaller than the frequency distance of 127~GHz separating the two modes.

The $N = 4$ system is examined in Fig.~\ref{fig:linewidth}(b). This time the neighbor of the toroidal mode is the quadrupole shown in Fig.~\ref{fig:fourrod}(a) (mode B). As already noted in Sect.~\ref{sec:level42} the two modes are not well separated for small $s$ values. In particular, for $s_\mathrm{min} = 17~\mu$m the spectral separation of the central resonances is only 35~GHz. Although the resonance linewidth of the second mode is narrow, exciting the toroidal mode alone would be challenging. For greater separation values the behavior of the system is similar to the $N = 3$ case with the two resonances overlapping for separation values up to $60~\mu$m. The spectral isolation increases only after the toroidal mode has entered the TE$_{10}$-to-TE$_{00}$ collective mode transition phase.

\begin{figure*}
\includegraphics{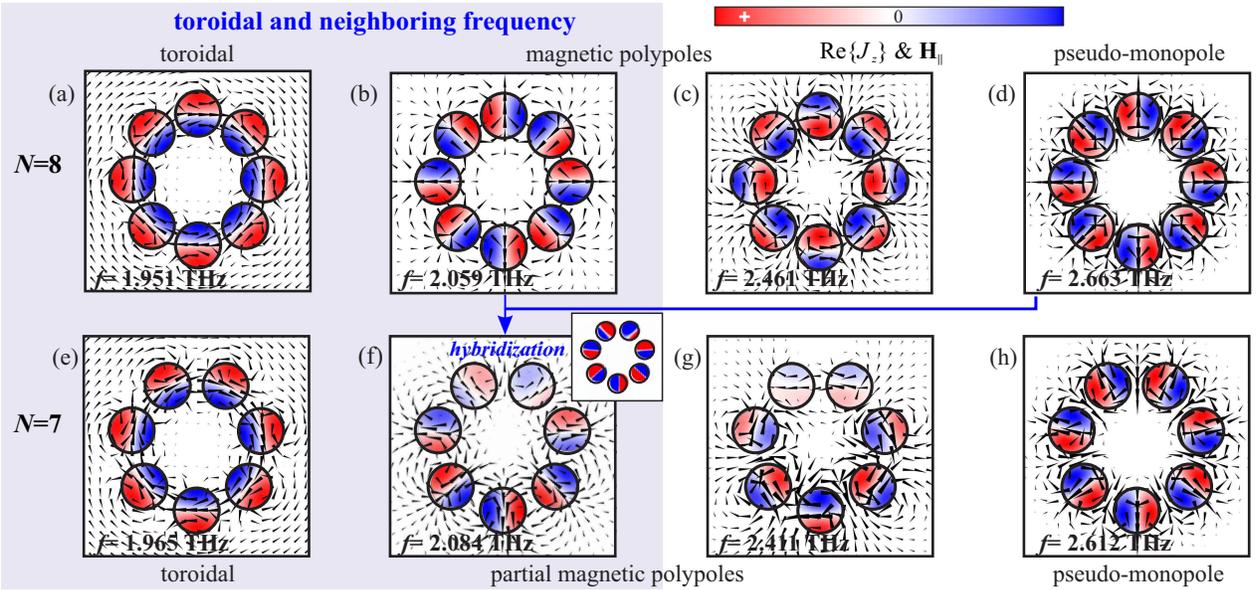}
\caption{\label{fig:78rods}Polarization current distribution for selected TE$_{10}$ collective modes  ordered in ascending frequency, for $N=8$ (a-d) and $N=7$ (e-h) cyclic metamolecules. The local magnetic dipoles are either azimuthally or radially aligned with fixed or alternating directions. Modes (a) and (e) are the toroidal modes of the $N=8$ and $N=7$ systems and (b) and (f) are the closest neighbors in frequency. In the $N=7$ system, mode (f) emerges as a hybridization of mode types (b) and (d), resulting in an increase of its resonant frequency.}
\end{figure*}

Moving on to systems with $N>4$ we find that spectral isolation is enhanced (at the cost of a larger metamolecule). Interestingly, the advantage of the $N=3$ compared to the $N=4$ system is generally observed when comparing $N=2n-1$ with $N=2n$ systems. In particular, we compare the first row in Fig.~\ref{fig:linewidth} examining odd-numbered systems with the second row of Fig.~\ref{fig:linewidth} examining even-numbered systems. Systems with $N=3,5,7$ exhibit systematically higher spectral isolation for the toroidal mode compared with $N=4,6,8$ systems: the frequency distances between the toroidal mode and its neighbor at $s_\mathrm{min}=17~\mu$m are $\{$127, 35, 180, 100, 193, 140$\}$~GHz for $N=3-8$, respectively. This behavior is attributed to the characteristics of the mode close to the toroidal. To better demonstrate this, we present in Fig.~\ref{fig:78rods} a comparison of selected TE$_{10}$ collective modes for the $N=7$ and $N=8$ systems. Figure~\ref{fig:78rods}(a-d) depicts four characteristic modes of the $N = 8$ system in ascending resonance frequency and for a rod separation of $s = 21~\mu$m. They are formed by an azimuthal or radial arrangement of the local magnetic dipoles with fixed or alternating directions. The corresponding modes of the $N = 7$ system are depicted in Fig.~\ref{fig:78rods}(e-h). Modes (a), (e) are the toroidal modes of the systems and modes (b), (f) their respective (partial) magnetic polypole neighbors. We are interested in determining why (e) is more separated from (f) than (a) from (b). Focusing on the $N=8$ case, we note that modes (b) and (d) are characterized by a radial arrangement of the local dipoles with alternating and fixed directions, respectively. In mode (b) (alternating local moment directions) the nearby currents in adjacent rods are of the same sign and the electric field experiences a variation with $N=8$ zeros along the fictitious circumference connecting the rod axes. On the other hand, in mode (d) the nearby currents in the adjacent rods are of opposite sign and the number of the zeros is $2N=16$, explaining the higher frequency of mode (d): 2.663 vs 2.059 THz. In the $N=7$ system, mode (f) (corresponding to mode type (b) of the 8-rod system) is not so well-defined. In particular, mode (f) fails to fulfill the type (b) distribution, since the alternating direction is not commensurate with the odd number of rods.
Instead, mode (f) emerges as a hybridization of mode type (b) with mode type (d) and the number of zeros is $N+1=7+1=8$ (the current distribution is shown saturated in the inset of Fig.~\ref{fig:78rods}(f) to better illustrate this fact). This results in an increase of its resonant frequency (recall that mode type (d) features a faster  variation along the circumference), leading to a larger frequency separation from the toroidal mode. This also explains why the advantage of odd-numbered systems diminishes as $N$ increases: the characteristics of mode type (d) are inherited for only one pair of adjacent rods meaning that the bump in frequency becomes less pronounced as $N$ increases. Note, finally, that the second partial magnetic polypole of the $N=7$ system, shown in Fig.~\ref{fig:78rods}(g), emerges in a similar manner as a hybridization of mode types (a) and (c) which results in a decrease of its resonant frequency.

\subsection{\label{sec:level45}Cyclic metamolecules of elliptical rods}

\begin{figure}
\includegraphics{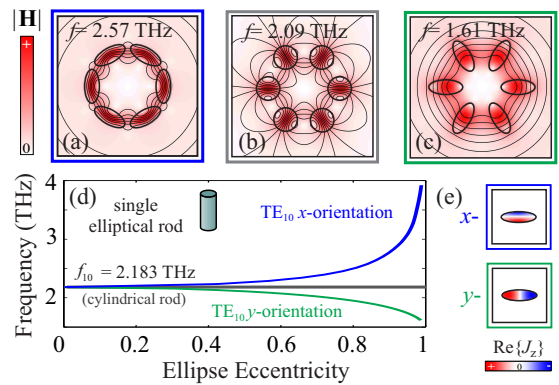}
\caption{\label{fig:ellipse} Toroidal mode: magnetic field lines and magnetic field $|\mathbf{H}|$ (color) for the six-rod system of (a)~elliptical cross-section arranged with the major axes along the azimuthal direction, (b)~circular cross-section and (c)~elliptical cross-section with the major axes along the radial direction. The eccentricity of the ellipse is equal to $e=0.9$ and the radius of the metamolecules is equal to $25~\mu$m. (d)~Dependence of the single-rod TE$_{10}$ resonant frequency on the eccentricity. Frequency splitting for the $x$-(blue curve) and the $y$-oriented (green curve) magnetic dipoles as shown in Fig. \ref{fig:ellipse}(e). (e)~Polarization current distribution of the $x$- and $y$-oriented magnetic dipoles.}
\end{figure}

In the expectation of a toroidal mode closer to the ideal one, we investigate finally the case of cyclic metamolecules made of elliptical rods of equal cross-sectional area to the circular ones. In particular, we investigate the toroidal mode supported by a six-elliptical-rod metamolecule and compare it with that of the circular-rod counterpart, Fig.~\ref{fig:ellipse}. The radius of the metamolecule is constant throughout and equal to $25~\mu$m and the eccentricity, $e$, of the elliptical rods in Fig.~\ref{fig:ellipse}(a,c) is equal to $e=0.9$. The magnetic field lines and magnetic field distribution (color) for the toroidal mode are presented for three distinctive cases. In Fig.~\ref{fig:ellipse}(a) the elliptical rods are arranged with their major axes along the azimuthal direction, in Fig.~\ref{fig:ellipse}(b) the rods are  of circular cross-section and in Fig.~\ref{fig:ellipse}(c) the elliptical rods are arranged with their major axes along the radial direction. Comparing the field distribution in  Fig.~\ref{fig:ellipse}(a) and Fig.~\ref{fig:ellipse}(b), we observe that the confinement of the magnetic field lines in the elliptical rods  with the azimuthal arrangement is enhanced; this implies the formation of better-defined toroidal modes. On the contrary, in the case of the radial arrangement, Fig.~\ref{fig:ellipse}(c), the magnetic field lines escape the characteristic path of the torus leading to poorly-defined toroidal modes. It is also interesting to note that the resonant frequency of the toroidal mode for the three systems is very different: 2.57,  2.09 and 1.61~THz for cases (a)-(c), respectively. This is due to the fact that the toroidal mode is formed by different single-rod TE$_{10}$ modes as a result of the cylindrical symmetry breaking lifting the degeneracy. Indeed, the ellipse supports two TE$_{10}$ magnetic dipole modes with different resonant frequencies: one with the magnetic moment along the major axis ($x$-orientation) and the other along the minor axis ($y$-orientation) [the distribution of the corresponding currents  for $e=0.9$ are depicted in Fig.~\ref{fig:ellipse}(e)]. Figure~\ref{fig:ellipse}(d) shows the evolution of the resonant frequencies for the $x$-oriented (blue curve) and $y$-oriented (green curve) TE$_{10}$ modes with respect to the ellipse eccentricity.
Actually, the $x$-oriented TE$_{10}$ mode is expected to be close but higher than the mode in a slab of thickness equal to the minor axis of the ellipse and length equal to the major axes of the ellipse. In particular for the case of eccentricity equal to $e=0.9$, the ratio between the $x$-oriented TE$_{10}$ resonance in the ellipse and the corresponding resonance of a slab with cross-section that encloses the ellipse (circumscribed rectangle) is 2.77~THz : 2.49~THz.
The toroidal of Fig.~\ref{fig:ellipse}(a) is formed by $x$-oriented magnetic dipoles which explains its high frequency; in analogy, the low frequency toroidal mode, Fig.~\ref{fig:ellipse}(c), is formed by $y$ dipoles. It is important to stress out here that the toroidal spectral isolation is compromised in the elliptical rods metamolecules; this is due to the frequency splitting of the isolated rod modes. Similarly to the TE$_{10}$ magnetic dipole in the ellipse, we also expect frequency splitting of the lower and higher order neighbor modes (e.g. of the type TE$_{00}$, TE$_{01}$ and TE$_{20}$ seen in Fig.~\ref{fig:singlerod}). This frequency splitting is expected to lead to a higher spectral overlap between the collective modes in the many-elliptical rod metamolecules.

\section{\label{sec:level5}Conclusion}
We have presented a thorough investigation of the electromagnetic resonant modes supported by cyclic metamolecules of $N=2$ to $8$ polaritonic rods. We have focused our study on TE$_{10}$-based collective modes since the TE$_{10}$ mode is the building block for the formation of the peculiar toroidal dipole. In each system, we have identified the toroidal mode (both by the distribution of the fields and the explicit calculation of the toroidal dipole moment as defined by Eq.~\ref{eq:tordipole}) and those lying in its spectral neighborhood and we have investigated the features of the resonances with varying rod separation. We have conducted the analysis with finite-element eigenvalue simulations and the results have been complemented with coupled mode theory and a lumped wire model capturing the coupling-caused reorganizations of the currents in each rod in analogy with the reorganization of the changes in each atom within the framework of the LCAO in molecular and solid state physics. We found that the collective mode eigenfrequencies oscillate about the single-rod magnetic dipole resonance, a feature attributed to the leaky nature of the constituent modes. We have also shown that metamolecules with an odd number of rods exhibit enhanced spectral isolation for the toroidal mode; along with its leaky nature this can lead to configurations that favor the unambiguous excitation and detection  of the unconventional toroidal response.

\section{\label{sec:level6}Acknowledgements}
This work was supported by the European Research Council under ERC Advanced Grant No. 320081 (PHOTOMETA). Work at Ames Laboratory was partially supported by the Department of Energy (Basic Energy Sciences, Division of Materials Sciences and Engineering) under Contract No. DE-AC02-07CH11358.
\appendix
\section{Coupled Mode Theory Framework}\label{sec:CMT}

The framework used is based on Refs.~\onlinecite{Haus:1991,Popovic:2006}. We briefly outline the formulation to highlight the points requiring attention. The modes of the $N-$rod structure can be specified by finding the extrema of the functional
\begin{equation}\label{eq:Functional}
  \omega^2=\frac{\displaystyle\int\nabla\times\mathbf{H}\cdot\widetilde{\varepsilon}^{-1}\nabla\times\mathbf{H}\,d\widetilde{\Omega}}%
  {\displaystyle\int\mathbf{H}\cdot\widetilde{\mu}\,\mathbf{H}\,d\widetilde{\Omega}},
\end{equation}
where $\widetilde{\varepsilon}$ and $\widetilde{\mu}$ are tensors for the general case of anisotropic materials. This functional is found by dot multiplying the $H$-field vector-wave equation with $\mathbf{H}$ (unconjugated to allow for leaky modes) and integrating. Using the magnetic field as the working variable is important as will be shown in the next paragraph. The form of Eq.~\eqref{eq:Functional} is reached only when the boundary term $\oint\mathbf{H}\cdot\left(\hat{\mathbf{n}}\times\widetilde{\varepsilon}^{-1}\nabla\times\mathbf{H}\right)\,d\Gamma$ which arises is zero. In open, leaky systems this is handled by surrounding the structure with perfectly matched layers (PMLs) backed with a PEC/PMC boundary condition. For stretched-coordinate PMLs, integration in Eq.~\eqref{eq:Functional} extends in the complex plane, denoted by $d\widetilde{\Omega}$. Including the PML in the integration domain also provides a means of compensating for the exponential divergence observed in the field profile of complex-frequency eigenmodes \cite{Sauvan:2013}.

Next, we suppose that the supermodes supported by the structure can be expressed as a linear combination of the $N$ isolated-rod modes $\mathbf{H}=\sum_{i=1}^N a_i\mathbf{H}_i$. In other words, we assume that coupling does not significantly perturb the individual modes comprising the supermode. Using the magnetic field in the expansion (and the corresponding version of the functional, Eq.~\eqref{eq:Functional}) is crucial in order for the supermode trial function to satisfy the divergence condition $\nabla\cdot\mathbf{D}=0$ \cite{Johnson:2002}. If the electric field is instead used then $\nabla\cdot\widetilde{\varepsilon}\mathbf{E}=\sum_{i=1}^N\nabla\cdot\widetilde{\delta\varepsilon}_i\mathbf{E}_i\neq0$, where $\widetilde{\varepsilon}\equiv\widetilde{\varepsilon}_i+\widetilde{\delta\varepsilon}_i$. Obviously, each mode $\mathbf{H}_i$ satisfies a vector wave equation of its own. Taking the inner product with $\mathbf{H}_j$ and omitting the boundary term (zero due to the use of PMLs) we can write
\begin{equation}\label{eq:VWEqHi}
  \int\nabla\times\mathbf{H}_j\cdot\widetilde{\varepsilon}_i^{-1}\nabla\times\mathbf{H}_i\,d\widetilde{\Omega}
=\omega_i^2\int\mathbf{H}_j\cdot\widetilde{\mu}_i\mathbf{H}_i\,d\widetilde{\Omega}.
\end{equation}

The linear combination is substituted in Eq.~\eqref{eq:Functional} which can be written in matrix form (uppercase italic bold symbols indicate $N\times N$ matrices, whereas lowercase italic bold symbols indicate $N\times 1$ vectors):
\begin{equation}\label{eq:FunctionalMatForm}
  \omega^2=\frac{\bm{a}^T\bm{K}\bm{a}}{\bm{a}^T\bm{W}\bm{a}},
\end{equation}
where
\begin{subequations}
\begin{align}
  \label{eq:Kmatrix}K_{ij}&\equiv\int\nabla\times\mathbf{H}_i\cdot\widetilde{\varepsilon}^{-1}\nabla\times\mathbf{H}_j\,d\widetilde{\Omega},\\
  \label{eq:Wmatrix}W_{ij}&\equiv\int\mathbf{H}_i\cdot\widetilde{\mu}\,\mathbf{H}_j\,d\widetilde{\Omega}.
\end{align}
\end{subequations}
Differentiating the right hand side of Eq.~\eqref{eq:FunctionalMatForm} with respect to the complex $a_i$ as shown in Ref.~\onlinecite{Haus:1991} we reach
\begin{equation}\label{eq:MasterEq}
\bm{Ka}-\omega^2\bm{Wa}=0.
\end{equation}
Importantly, the supermode frequencies can be directly determined form solving Eq.~\eqref{eq:MasterEq} which amounts to finding the eigenvalues of the $\bm{W}^{-1}\bm{K}$ matrix (and taking the square root). There is no need to first derive a temporal CMT equation (see Refs.~\onlinecite{Haus:1991,Popovic:2006}), something which entails the assumption that all supermode frequencies cluster around a typical value $\omega_0$ making it more approximate. Note that for evaluating the $\bm{K}$ matrix it is necessary to make use of Eq.~\eqref{eq:VWEqHi}. This permits us to incorporate in the formulation the zeroing of the boundary term for each constituent mode, as we did for the entire supermode in the process of reaching Eq.~\eqref{eq:Functional}. To this end, we introduce in Eq.~\eqref{eq:Kmatrix} the perturbations to individual permittivity distributions $\widetilde{\varepsilon}^{-1}\equiv\widetilde{\varepsilon}_j^{-1}+\widetilde{\Delta\varepsilon}_j^{-1}$, and use Eq.~\eqref{eq:VWEqHi} to arrive at
\begin{equation}\label{eq:Kmatrix2}
\begin{split}
K_{ij}&=\omega_j^2\int\mathbf{H}_i\cdot\widetilde{\mu}_j\mathbf{H}_jd\widetilde{\Omega}-
\omega_i\omega_j\int\mathbf{D}_i\cdot\widetilde{\Delta\varepsilon}_j^{-1}\mathbf{D}_jd\widetilde{\Omega},\\
&=W_{ij}\omega_j^2-2\omega_iM_{ij}\omega_j.
\end{split}
\end{equation}
Matrix $\bm{M}$ in Eq.~\eqref{eq:Kmatrix2} is responsible for mode coupling. Note that integration is restricted to regions where $\widetilde{\Delta\varepsilon}_j^{-1}$ is nonzero. The off-diagonal elements ($i\neq j$) describe resonator-to-resonator coupling (frequency splitting), whereas the elements on the main diagonal ($i=j$) represent CIFS \cite{Popovic:2006}, i.e., the modification of the isolated-rod frequencies due to the index perturbations experienced by their own field profiles. In the case of evanescent coupling, the $M_{ij}$ elements monotonically decay with resonator separation; it is only when oscillating tails are involved that they oscillate between positive and negative values. This oscillating behavior is inherited by the supermode frequencies as can be seen by writing
\begin{equation}\label{eq:WKproduct}
\begin{split}
\bm{W}^{-1}\bm{K}&=\left(\bm{\Omega}_d-2\bm{W}^{-1}\bm{\Omega}_d\bm{M}\right)\bm{\Omega}_d\\
&=\left(\bm{\Omega}_d-2\bm{\Lambda}\right)\bm{\Omega}_d,
\end{split}
\end{equation}
where we have introduced the diagonal matrix $\bm{\Omega}_d=\operatorname{diag}(\omega_1,\omega_2, \ldots, \omega_N)$ containing the isolated-rod resonant frequencies and defined $\bm{\Lambda}\equiv\bm{W}^{-1}\bm{\Omega}_d\bm{M}$ which has units of frequency (both $\bm{W}$ and $\bm{M}$ are measured in Joules).

Note that if we further assume that $\bm{\Lambda}$ represents a small perturbation to the resonant frequency, then $\bm{\Lambda}^2$ is of second-order smallness. Therefore, in the context of first-order perturbation theory we can write $\left(\bm{\Omega}_d-2\bm{\Lambda}\right)\bm{\Omega}_d\approx\left(\bm{\Omega}_d-\bm{\Lambda}\right)^2$, completing the binomial identity, and recover the result in Ref.~\onlinecite{Popovic:2006} which states that one can solve for the supermode frequencies (instead of their squares) by finding the eigenvalues of the $\bm{\Omega}_d-\bm{\Lambda}$ matrix.

\section{Lumped Wire Model}\label{sec:WM}
\begin{figure}
\includegraphics{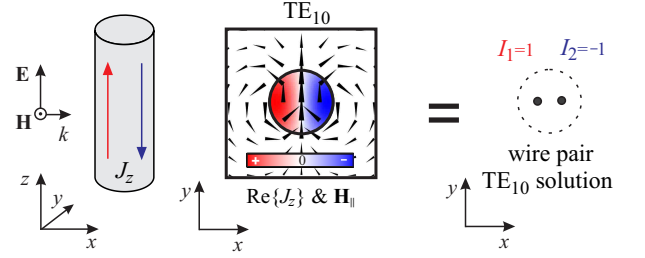}
\caption{\label{fig:wmappendix}TE$_{10}$ magnetic dipole mode: Schematic for the description of the lumped wire pair equivalent.}
\end{figure}

In the lumped wire model approximation we assume that the features of each constituent TE$_{10}$ magnetic dipole mode can be approximated by a combination of thin wires,  infinitely long along the $z$ direction, that carry uniform currents. As seen in Fig.~\ref{fig:wmappendix}, the TE$_{10}$ dipole mode is characterized by two separated symmetric areas of positive and negative oscillating displacement currents. Assuming the simplest possible approximation, we consider that a pair of wires with currents $I_1=1$ and $I_2=-1$, placed at a fixed position is able to reproduce the features of the mode. We note here that the choice of the two wires facilitates the simplicity of the model; a more accurate representation of the TE$_{10}$ would occur by a combination of a larger number of current-carrying wires. For the $N$ rods system and for the TE$_{10}$-based collective modes we assume $N$ pairs of wires placed at the desirable separation distance. We expect that in each wire the currents should be capable of reproducing locally TE$_{10}$-like field distributions. Up to now we have formed a system of $M = 2N$ coupled current-carrying wires; each current $I_m$ radiates omnidirectional electromagnetic energy and at the same time the current in each wire is affected by the radiation coming from the adjacent $M-1$ wires.
The radiation field that each wire $m$ produces and in particular the $E_z$ component of the electric field, reads
\begin{equation}\label{radiationIm}
E_{z,m}(\mathbf{r}) = -I_m \frac{\mu_0}{4} \omega H_0^{(1)} (k_0|\mathbf{r}-\mathbf{r_m}|),
\end{equation}
where $|\mathbf{r}-\mathbf{r_m}|$  is the distance from the $m^\mathrm{th}$ wire in the $xy$ plane. At the position of each wire the total field coming from the adjacent $M-1$ wires is the sum of each $M-1$ radiation contribution and at the same time it is equal to the local electric field produced by the current $I_m$, $E_{z,m}(\mathbf{r}\equiv\mathbf{r_m}) = I_m Z_m $, where $Z_m$  is a term that contains all the impedance contributions in the wire and it is the same for all  wires $Z_m=Z$, $\forall m$. For the electric field at each $m^\mathrm{th}$ wire is we have
\begin{equation}\label{radiationatIm}
I_m Z = \sum_{n \neq m}^{M}-I_n \frac{\mu_0 }{4} \omega H_0^{(1)} (k_0 |\mathbf{r_n}-\mathbf{r_m}|),
\end{equation}
where $|\mathbf{r_n}-\mathbf{r_m}|$  is the distance between the $m^{th}$  and the $n^{th}$ in the $xy$ plane.
The distances and the frequency is constant, $\omega = \omega_{10}$, and the free parameters are the currents $I_m$ and impedance terms $Z_m$. The system of the linear equations corresponds to an $M \times M$ eigenvalue problem, $\mathbf{L}\mathbf{I^{crt}}-Z\mathbf{I^{crt}}=0$, where
\begin{equation}\label{radiationatIm}
L_{nm} = -\sum_{n \neq m}^{M}\frac{\mu_0 }{4} \omega H_0^{(1)} (k_0 |\mathbf{r_n}-\mathbf{r_m}|).
\end{equation}

The system has $M$ eigenvalues $Z$   and eigenvectors $\mathbf{I^{crt}}=[I_1,I_2,...,I_{M-1},I_M]$   that correspond to the currents flowing through each wire.  Among the $M$ solutions of the problem we find the eigenvectors with currents that correspond to the TE$_{10}$-based collective mode under consideration. For example at the case for the isolated-single rod we place a pair of wires at fixed points at a distance $w<2R$, $R$ is the radius of the rods. The current carrying pair produces a $2 \times 2$ system with two solutions, with eigenvalues $Z_1$   and $Z_2$ , and eigenvectors  $\mathbf{I^{crt}}=[1,-1]$ and  $\mathbf{I^{crt}}=[1,1]$. Solution   $Z_1$ and $\mathbf{I^{crt}}=[1,-1]$  corresponds to the TE$_{10}$ dipole mode. We note here that $w$ is a parameter that can be finer tuned in order to approximate more effectively the corresponding mode and here is chosen equal to $7~\mu$m.

\end{document}